\documentclass[prd,aps,twocolumn,nofootinbib,superscriptaddress,eqsecnum,floatfix,preprintnumbers,amsmath,amssymb,
nofootinbib,longbibliography]{revtex4-1}

\usepackage{amssymb,amsmath}
\usepackage{epsfig}
\usepackage[dvipsnames]{xcolor}
\usepackage[utf8]{inputenc}
\usepackage{stmaryrd}
\usepackage{mathrsfs}
\usepackage{mathalfa}
\usepackage{accents}
\usepackage[normalem]{ulem}
\usepackage{enumitem}

\usepackage{graphicx}

\newcommand{\JJ}{\mathcal{J}}
\newcommand{\PP}{\mathcal{P}}
\newcommand{\EE}{\mathcal{E}}

\newcommand{\pa}{\partial}

\newcommand{\be}{\begin{equation}}
\newcommand{\ee}{\end{equation}}
\newcommand{\BS}{\begin{split}}
\newcommand{\ES}{\end{split}}
\newcommand{\bea}{\begin{eqnarray}}
\newcommand{\eea}{\end{eqnarray}}
\newcommand{\ba}{\begin{equation}\begin{aligned}}
\newcommand{\ea}{\end{aligned}\end{equation}}

\newcommand{\beg}{\begin{gather*}}
\newcommand{\eng}{\end{gather*}}
\newcommand{\hh}{,\hspace{0.5cm}}
\newcommand{\hhh}{,\hspace{0.2cm}}

\newcommand{\lap}{\triangle}

\newcommand{\n}[1]{\label{#1}}

\newcommand{\eqH}{\stackrel{H}{=}}

\newcommand{\CAL}{\mathcal}

\newcommand{\ts}[1]{{\boldsymbol{#1}}}

\usepackage[makeroom]{cancel}
\usepackage[caption=false]{subfig}
\usepackage[colorlinks=true,
            citecolor=green,
            linkcolor=red,
            filecolor=cyan,
            urlcolor=magenta,
            backref=false]{hyperref}

\newcommand{\Dr}{\Delta_r^{(0)}}
\newcommand{\Dy}{\Delta_y^{(0)}}

\begin{document}

\title{Motion of a rotating black hole in a homogeneous scalar field}

\author{Valeri P. Frolov}%
\email[]{vfrolov@ualberta.ca}
\affiliation{Theoretical Physics Institute, Department of Physics,
University of Alberta,\\
Edmonton, Alberta, T6G 2E1, Canada
}

\begin{abstract}
In the present paper, we consider a rotating black hole moving in a homogeneous massless scalar field. We assume that the field is weak and neglect its backreaction, so that the metric at far distance from the black hole is practically flat. In this domain one can introduce two reference frames, $K$ and $\tilde{K}$. The frame $\tilde{K}$ is  associated with the homogeneous scalar field, in which its constant gradient has only time component. The other frame, $K$, is the frame in which the black hole is at rest. To describe the Kerr metric of the black hole we use its  Kerr-Schild form $g_{\mu\nu}=\eta_{\mu\nu}+\Phi l_{\mu}l_{\mu}$, where $\eta_{\mu\nu}$ is the (asymptotic) flat metric in $K$ frame. We find an explicit solution of the scalar field equation which is regular at the horizon and properly reproduce the asymptotic form of the scalar field at the infinity. Using this solution we calculate the fluxes of the energy, momentum and the angular momentum of the scalar field into the black hole. This allows us to derive the equation of motion of the rotating black hole. We discuss main general properties of solutions of these equations and obtain explicit solutions for special type of the motion of the black hole.

\hfill {\scriptsize Alberta Thy 9-23}
\end{abstract}


\maketitle

\section{Introduction}

Scalar field plays an immense role in the modern physics and cosmology. In the high energy physics, a scalar Higgs field is used to provide the particles their mass as a result of the spontaneous symmetry breaking. In cosmology, the inflation can be driven by the potential part of the scalar field ("inflaton"). The scalar field and Higgs mechanism are important parts of the models describing possible symmetry breaking and phase transitions in the cosmology. It is believed that in the early Universe  there might be several of such phase transitions which played an important role in its evolution. During these transitions formation of primordial black holes and cosmic strings might become possible. More recently, another mechanism of symmetry breaking known as the ghost condensation was proposed  \cite{Arkani-Hamed:2003pdi,Arkani-Hamed:2003juy}.
The corresponding phase of the ghost condensate is formed due to the special form of the kinetic part of the scalar field action. In such a model the ground state is the scalar field with nonvanishing vector of its gradient. The presence of such a field breaks the Lorentz invariance.

The ghost condensation model belongs to a wide class of scalar field models which are invariant with respect to the scalar field shift $\Psi\to\Psi+\mbox{const}$. At the lowest order in derivatives the Poincar\'e invariant Lagrangian for such a theory takes the form
\be
L=L(X)\hh X=\eta^{\mu\nu}\Psi_{,\mu}\Psi_{,\nu}\, .
\ee
Such a model can also be used as the low-energy effective field theory for zero- and finite-temperature relativistic superfluids \cite{Son:2002zn,Nicolis:2011cs}. For the superfluid state with finite charge density and vanishing spatial current the corresponding solution is $\Psi=\mu t$, where $\mu$ is the chemical potential \cite{Son:2002zn}.
Similar solutions with a constant spacelike gradient of the scalar field were considered in application to the cosmology in the framework of a so-called solid inflation model \cite{Endlich:2012pz}.
Let us also mention an interesting approach in which the scalar field is used to describe a phenomenon emergence of time and dynamics in originally Euclidean spacetime (see e.g. \cite{Moffat:1992bf,Babichev:2007dw,Mukohyama:2013ew,Mukohyama:2013gra}).

A natural and interesting question is how a black hole interacts with a scalar field in different models and under different conditions. Let us note that   black holes cannot have their own scalar field. Discussion of the
 no-hair theorem for the scalar field and further references can be found in \cite{Bekenstein:1995un}. At the same time a black hole can exist in the presence of an external scalar field. Accretion of the ghost condensate by black holes in the expanding universe was discussed in  \cite{Frolov:2002va,Frolov:2004vm,Mukohyama:2005rw,Barranco:2011eyw}.
 Primordial black holes and Higgs field vacuum decay were considered in \cite{Gregory:2013hja,Chadburn:2013mta,Gregory:2018ghc,Gregory:2023eos}. Scalar field accretion by black holes was also discussed in \cite{deCesare:2022aoe}.

 In this paper we consider a motion of a rotating black hole in an external homogeneous massless minimally coupled scalar field. This model allows a rather complete analysis. In the absence of the black hole, the scalar field equation in the flat spacetime
 \be \n{EQB}
 \Box \Psi=0\, ,
 \ee
 has a simple solution
 \be \n{AYM}
 \Psi=\Psi_0 t\, .
 \ee
 Such a field has constant gradient and is homogeneous in space. This form of the solution is valid in a specially chosen inertial reference frame, and, in this sense, it breaks the Lorentz invariance.
 This choice of the solution is motivated by the ghost condensate model.
 We consider a rotating black hole moving in such a scalar field. In the presence of the black hole the scalar field $\Psi$ is distorted. We assume that the scalar field is weak and its backreaction on the metric can be neglected.

 Our first goal is to find a solution for the scalar field which is regular at the horizon of the moving black hole and has the asymptotic form \eqref{AYM} at far distance from it. We shall demonstrate that this problem allows an exact solution. To find this solution we proceed as follows.

A remarkable property of the Kerr metric is that it can be written in the Kerr-Schild  form \cite{Kerr_Schild}
\begin{equation}\n{KS}
g_{\mu\nu}=\eta_{\mu\nu}+\Phi l_{\mu}l_{\nu}\, ,
\end{equation}
where $\eta_{\mu\nu}$ is a flat metric, $\Phi$ is a scalar field, and $\ts{l}$ is a  tangent vector to a shear-free geodesic null congruence. It has been shown that these solutions of the Einstein equations can be obtained by complex coordinate transformations from the Schwarzschild metric \cite{Newman:1965tw,Newman:1973afx}. In particular, the potential $\Phi$ for the Kerr metric can be obtained as a solution of the Laplace equation in flat coordinates $(X,Y,Z)$
\begin{equation}\n{lap}
\lap\Phi=4\pi j\, ,
\end{equation}
with a point-like source $j$ located at the complex coordinate $Z+ia$, where $a$ is the rotation parameter of the Kerr black hole \cite{Israel,Kaiser_2003}. A comprehensive review of the Kerr-Schild metrics and complex space approaches can be found in \cite{Adamo:2014baa}. More recently, the Kerr-Newman representation of the spacetime geometry attracted a lot of attention in the so-called double copy formalism. This formalism is  based on the following result: Einstein equations for the metrics which allow the Kerr-Schild representation can be reduced to the linear equations for Maxwell and scalar fields. At the moment there exist dozens of publications on this subject. Related references can be found e.g. in the following review articles \cite{White_2018,bern2019duality,bern2022sagex}.

One can interpret this result as follows. The form \eqref{KS} of the metric allows one to treat the metric $\ts{\eta}$ as the metric of the background  flat spacetime, while the term $\Phi l_{\mu}l_{\nu}$ describes its  "perturbation" due to the black hole located at the origin of the background space. Denote by $M$ the mass of the black hole. The gravitational field of the black hole is strong in its vicinity. For an observer located at the distance $L\gg M$ from the black hole this field is weak. One can say that such an observer "lives" in the space with the background metric $\ts{\eta}$. Such an observer can describe the black hole as a small compact object and use for the description of its motion a "point particle" approximation.

If the scalar field is present and has the form \eqref{AYM}  the interaction of this field results in its accretion by the black hole. In this paper we consider a  black hole moving in such a homogenous scalar field. In this case their exist two natural reference frames. One of them, which we denote by $\tilde{K}$, is the frame in which at far distance from the black hole the scalar field has the form \eqref{AYM}. The other frame, moving with respect to $\tilde{K}$ with the velocity $\vec{V}$, is the frame in which the black hole is at rest. We denote it by $K$. In this frame the form of the solution for the scalar field differs from \eqref{AYM} and it can be obtained by making the correspondent Lorentz transformation. We shall use both frames. Namely, we use the Kerr-Schild form of the Kerr metric associated with $K$ frame to solve the scalar field equations in the presence of the black hole. Using this result we calculate the force acting on the black hole due to its motion in the scalar field and obtain the equation of motion of the black hole in $\tilde{K}$ frame.

The paper is organized as follows. In section~II we collect useful formulas connected with the Kerr-Schild form of the Kerr metric. Section~III describes a solution for the scalar field in the presence of the moving rotating black hole. Fluxes of the scalar field in $K$ frame are obtained in section~IV. Section~V  contains calculation of the 4D force acting on the black hole in the frame $\tilde{K}$. It also discusses the equations of motion of the black hole, general properties of their solutions, and special cases. Useful information concerning complex null tetrads is collected in the appendix~A. The appendix~B contains details of the calculations of the fluxes of the energy and angular momentum of the scalar field through the horizon of the black hole.
In the paper we use units in which $G=c=1$ and sign conventions  of the book \cite{MTW}.

\section{Kerr metric and its Kerr-Schild form}
\n{S2}
\subsection{The Kerr metric}

The Kerr metric, describing a vacuum stationary rotating black hole, written in the Boyer-Lindquist coordinates is
\begin{equation}\label{Kerr}
\begin{split}
ds^2 =& -\left( 1-\frac{2Mr}{\Sigma}\right) dt^2
-\frac{4Mar\sin^2\theta}{\Sigma} dt d\varphi\\
+&\left(r^2+a^2+\frac{2M a^2 r}{\Sigma} \sin^2\theta\right)\sin^2\theta d\varphi^2\\
+&\frac{\Sigma}{\Delta} dr^2 +\Sigma d\theta^2\, ,\\
&\Sigma=r^2+a^2\cos^2\theta\hh \Delta=r^2-2Mr+a^2 \,  .
\end{split}
\end{equation}
Here $M$ is the black hole mass, and $a$ is its rotation parameter. This metric has two commuting Killing vectors $\ts{\xi}=\pa_t$ and $\ts{\zeta}=\pa_{\phi}$. Let us denote
\be
r_{\pm}=M\pm b\hh b=\sqrt{M^2-a^2}\, .
\ee
Equation $r=r_+$, where $\Delta=0$, describes the event horizon. The surface area of the horizon is
\be
\CAL{A}=4\pi(r_+^2+a^2)=8\pi M r_+ \, .
\ee

Coordinates $(t,r,\theta,\varphi)$ are singular at this surface. To describe both the exterior and interior of a rotating black hole one can use so called Kerr incoming coordinates $(v,r,\theta,\tilde{\varphi})$ which are regular at the future event horizon \cite{Kerr}
\ba
&dv=dt+dr_*\hh dr_*=(r^2+a^2)\dfrac{dr}{\Delta}\, ,\\
&d\tilde{\varphi}=d\varphi +a\dfrac{dr}{\Delta}\, .
\ea
In these coordinates the metric \eqref{Kerr} takes the form
\ba
&ds^2=-\dfrac{\Delta}{\Sigma}\left( dv-\dfrac{1}{a}\Dy\, d\tilde{\varphi}\right)^2+
\dfrac{\Dy}{\Sigma}\left( dv-\dfrac{1}{a}\Dr\, d\tilde{\varphi}\right)^2\nonumber\\
&+\dfrac{\Sigma}{\Dy} dy^2+2 dr\, \left( dv-\dfrac{1}{a}\Dy\, d\tilde{\varphi}\right)\, .
\ea

Similarly, one can introduce Kerr outgoing coordinates $(u,r,\theta,\tilde{\varphi})$
\be
dv=dt-dr_*\hh d\tilde{\varphi}=d\varphi -a\dfrac{dr}{\Delta}\, ,
\ee
which are regular at the past horizon and cover the white hole domain.

\subsection{Useful coordinates}

For $M=0$ the Riemann curvature of the Kerr metric vanishes and the metric (\ref{Kerr}) becomes flat. We write it  the form\footnote{Let us note that we use notations $T$ and $\phi$ for the time and angle variables in the flat spacetime. These coordinates are used in the Kerr-Schild form of the Kerr metric while the standard Bouer-Lindquist coordinates $t$ and $\varphi$ are related to $T$ and $\phi$ by means of relations \eqref{KST}.}
\begin{equation}\label{Flat}
\begin{split}
dS^2&= - d{T}^2+dh^2\, ,\\
dh^2&=
\frac{\Sigma}{r^2+a^2} dr^2 +\Sigma d\theta^2+(r^2+a^2) \sin^2\theta d\phi^2 \, .
\end{split}
\end{equation}
The coordinates $(r,\theta,\phi)$ are oblate spheroidal coordinates taking the following values $r\ge 0$, $\theta\in [0,\pi]$, $\phi\in [0,2\pi]$. These coordinates are related to the Cartesian coordinates as follows
\begin{equation}
  \begin{split}
    X &= \sqrt{r^2 + a^2}\sin\theta \cos\phi \, , \\
    Y &= \sqrt{r^2 + a^2}\sin \theta \sin \phi  \, ,\\
    Z &= r \cos\theta \, .
  \end{split}
\end{equation}
In these coordinates the flat metric $dS^2$ takes the standard form
\be\n{CART}
dS^2=\eta_{\mu\nu}dX^2\, dX^{\nu}=-dT^2+dX^2+dY^2+dZ^2\, .
\ee

\begin{figure}[!hbt]
    \centering
      \includegraphics[width=0.4\textwidth]{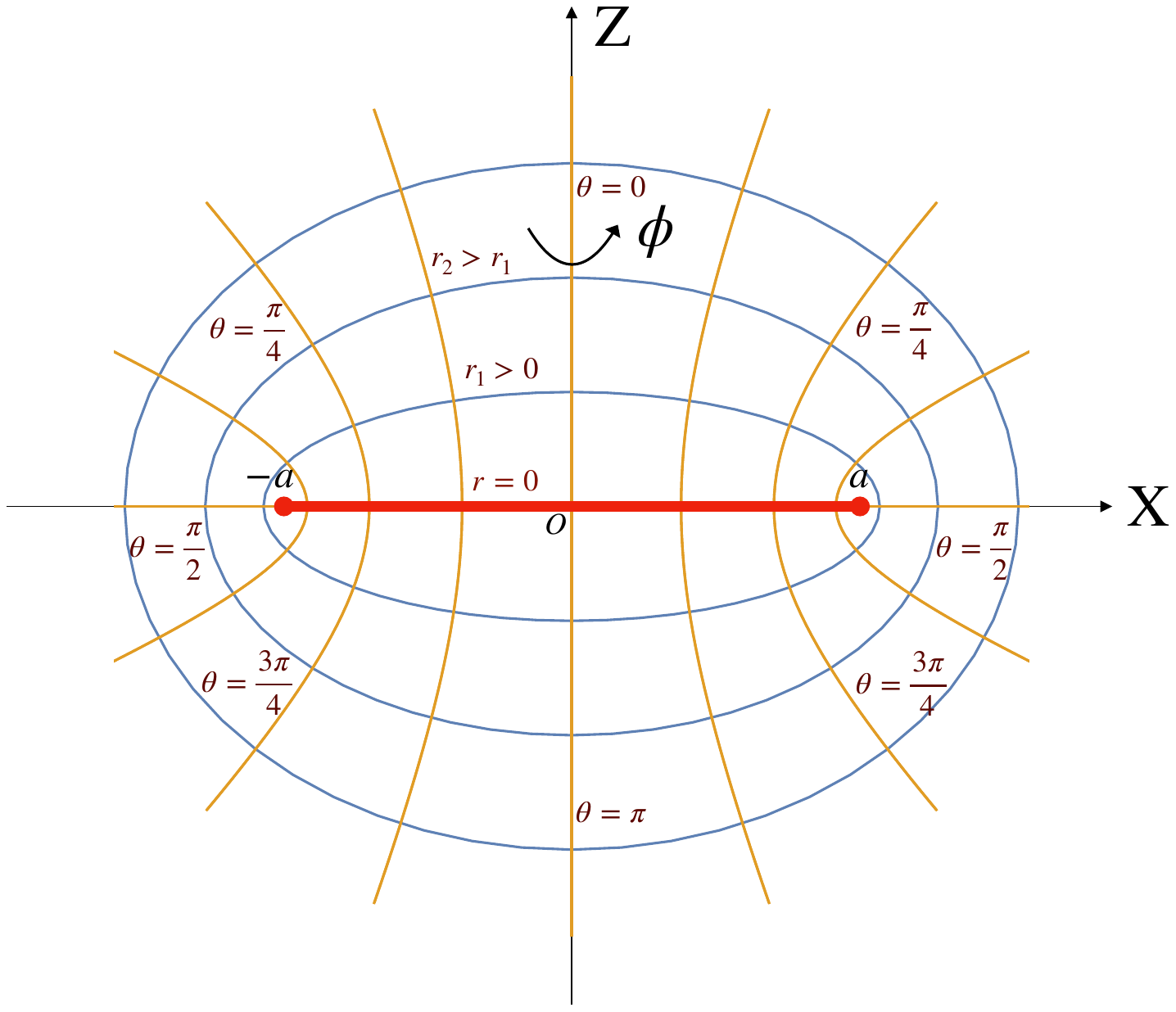}
   \caption{\n{F1} Coordinate lines of the oblate spheroidal coordinates $(r,\theta)$ in the plane $Y=0$}
\end{figure}

For $r>0$ the surfaces $r=$const are oblate ellipsoids.
Figure~\ref{F1} shows the coordinate lines of the oblate spheroidal coordinates $(r,\theta)$ in the plane $Y=0$ ($\phi=0$).
For $r=0$ and $\theta\in [0,\pi]$, $\phi\in [0,2\pi]$ one has a disc of radius $a$ located in the $Z=0$ plane. The coordinate $\theta$ is discontinuous on the disc. For $(0,\pi/2)$ the coordinate $\theta$ covers the upper part of the disc, while for  $(\pi/2,\pi)$, it covers the lower part of it.
The boundary  of this disc  is a ring of radius $a$.
Equations $\theta=0$ and $\theta=\pi$ describe the axis of symmetry $X=Y=0$. For $\theta=0$ and $Z=r$ it is positive, while for $\theta=\pi$ $Z=-r$ it is negative.

In what follows we shall also use another coordinate, $y$, related to the angle $\theta$ as follows
\begin{equation}
y=a\cos\theta.
\end{equation}

The flat metric $dS^2$ in the spheroidal coordinates $(T,r,y,\phi)$ is
\begin{equation}\n{RYMET}
\begin{split}
dS^2&=-dT^2+\Sigma\left( \dfrac{dr^2}{\Delta^0_r}+\dfrac{dy^2}{\Delta^0_y}\right)+\dfrac{\Delta^0_r\Delta^0_y}{a^2}d\phi^2\, ,\\
\Sigma&=r^2+y^2\hhh \Delta^0_r=r^2+a^2\hhh \Delta^0_y=a^2-y^2\, .
\end{split}
\end{equation}
In these coordinates the Cartesian coordinates take the form
\begin{equation}\n{XYZ_ry}
\begin{split}
X&=\dfrac{1}{a} \sqrt{\Dr\Dy}\cos\phi\, ,\\
Y&=\dfrac{1}{a} \sqrt{\Dr\Dy}\sin\phi\, ,\\
Z&=\dfrac{1}{a} ry \,  .
\end{split}
\end{equation}
We denote
\be\n{eee}
e_{(X)\mu}=X_{,\mu}\hh e_{(Y)\mu}=Y_{,\mu}\hh e_{(Z)\mu}=Z_{,\mu}\, .
\ee
We also denote by $\ts{e}_{i}$, $i=1,2,3$ a set of   4D unit vectors $e_i^{\mu}$  along $X$, $Y$ and $Z$ axes.

\subsection{Kerr metric in the Kerr-Schild form}
\n{S2}

Let us consider the following 1-form
\begin{equation}\n{lll}
l_{\mu}dx^{\mu}=-dT - \frac{\Sigma}{\Delta^0_r}dr+\dfrac{\Delta^0_y}{a}d\phi\, .
\end{equation}

We define a metric
\begin{equation}\n{KSP}
d{s}^2=dS^2+\Phi(l_{\mu}dx^{\mu})^2\, ,
\end{equation}
where $\Phi=\Phi(r,y)$ is some function.
The  metric coefficients of the metrics $ds^2$ and $dS^2$ are related as follows
\be\n{KSG}
g_{\mu\nu}=\eta_{\mu\nu}+\Phi l_{\mu} l_{\nu}\, .
\ee

The following statements are valid for each of the metrics $d{s}^2$ and $dS^2$. In other words, these statements are valid for an arbitrary function $\Phi$, including $\Phi=0$.
The vector field $\ts{l}$ has the following properties
\begin{itemize}
\item The contravariant components of the vector $\ts{l}$ in $(T,r,\theta,\phi)$ coordinates  are
    \be \n{LL}
    l^{\mu}=\left(1,-1,0,\dfrac{a}{r^2+a^2}\right)\, .
    \ee
\item $\ts{l}$ is a null vector $\ts{l}^2=l_{\mu}l^{\mu}=0$;
\item Vectors $\ts{l}$ are tangent vectors to incoming null geodesics in the affine parametrization,   $l^{\nu}l^{\mu}_{\ ;\nu}=0$.
\item $l^{\mu}_{\ ;\mu}=-\dfrac{2r}{\Sigma}$;
\item $l_{(\mu ;\nu)}l^{(\mu ;\nu)}-\frac{1}{2}(l^{\mu}_{\ ;\mu})^2=0$\, .
\end{itemize}

The last property implies that the congruence of null vectors $\ts{l}$ is shear-free (for more details see e.g. \cite{sommers1976,Frolov1979}).
Such a null geodesic congruence is related to the light cones with apex on the world-line in the complex space. The twist is a measure of how far the complex world-line is from the real slice \cite{Newman_2004}.

It is easy to check that for a special choice of the function $\Phi$
\begin{equation} \n{Phi0}
\Phi_0=\frac{2Mr}{\Sigma} \, ,
\end{equation}
the metric $d{s}^2$ given by \eqref{KSP} is Ricci flat, and in fact, it coincides with the Kerr metric. In order to prove this it is sufficient to make the following coordinate transformation
\begin{equation}
\begin{split}\n{KST}
&T = t +t_0(r)\hh \phi= \varphi  +\varphi_0(r)\, ,\\
&t_0(r)=\int\frac{2Mr}{\Delta}dr\\
&=\dfrac{M}{\sqrt{M^2-a^2}}\left[r_+\ln(r-r_+)-r_-\ln(r-r_-)\right]\, ,\\
&\varphi_0(r)=\int\frac{2Mar}{(r^2+a^2)\Delta} dr\\
&= \dfrac{a}{2\sqrt{M^2-a^2}}\ln\left( \dfrac{r-r+}{r-r_-}\right)-\arctan(r/a)+\dfrac{1}{2}\pi\, .
\end{split}
\end{equation}
Here $\Delta$ is defined in (\ref{Kerr}). These coordinates $(t,r,\theta,\varphi)$ are chosen so that the non-diagonal components
$g_{rt}$  and $g_{r\varphi}$ of the metric $d{s}^2$ vanish. One can check that the metric $d{s}^2$ written in the $(t,r,\theta,\varphi)$ coincides with the Kerr metric $dS^2$, provided one identifies the coordinates $t$ and $\varphi$ in $d{s}^2$ with the standard Boyer-Lindquist coordinates $t$ and $\varphi$ in the metric (\ref{Kerr}).
The integration constant in the expression for $\varphi_0$ is chosen so that this quantity vanishes when $r\to\infty$. Hence in this limit the angle variables $\varphi$ and $\phi$ coincide.

It is easy to check that  the coordinates $T$ and $\phi$ are related to the Kerr incoming coordinates $v$ and $\tilde{\varphi}$ as follows
\be
T=v-r\hh \phi=\tilde{\varphi}-\arctan(r/a)\, .
\ee
Coordinates $(T,r,y,\phi)$ are regular at the future event horizon and
cover the  exterior region of the black hole as well as a part of its interior.  Similarly, by a simple change of the sign of the coefficient of $dr$ term in the  expression \eqref{lll} one can obtain outgoing null vector and use it to construct the Kerr-Schild metric which is regular at the past horizon.

\section{Scalar field}

\subsection{Flat spacetime}

Let us consider a minimally coupled massless scalar field $\Psi$ which obeys the equation
\be \n{BOX}
\Box\Psi=0\,
\ee
Its stress-energy tensor is
\be \n{TEN}
T_{\mu\nu}=\Psi_{,\mu}\Psi_{,\nu}-\dfrac{1}{2} g_{\mu\nu}\Psi_{,\alpha}\Psi^{,\alpha}\, .
\ee

Let us consider first the flat spacetime.  We choose some inertial frame in it.  We denote it by  $\tilde{K}$ and
denote by $\tilde{X}^{\mu}=(\tilde{T},{X},Y,\tilde{Z})$ the Cartesian coordinates associated with this frame.
In this frame there exists a simple solution of the \eqref{BOX}
\be \n{PKP}
\Psi=\Psi_0 \tilde{T}\, ,
\ee
describing a homogeneous scalar field. The stress-energy tensor for this solution is diagonal and has the following non-vanishing components
\be
T_{\tilde{T}\tilde{T}}=T_{\tilde{X}\tilde{X}}=T_{\tilde{Y}\tilde{Y}}=T_{\tilde{Z}\tilde{Z}}=\dfrac{1}{2}\Psi_0^2\, .
\ee

The gradient of the field $\Psi$ has only time component.
In this sense this solution breaks the Lorentz invariance and singles out an inertial
frame in which the field $\Psi$ does not depend on spatial coordinates. As we already mentioned such solutions with a constant gradient of the field play an important role the shift-invariant scalar field theories (see e.g.
\cite{Arkani-Hamed:2003pdi,Arkani-Hamed:2003juy,Nicolis:2011cs,Endlich:2012pz}. )

\begin{figure}[!hbt]
    \centering
\includegraphics[width=0.45\textwidth]{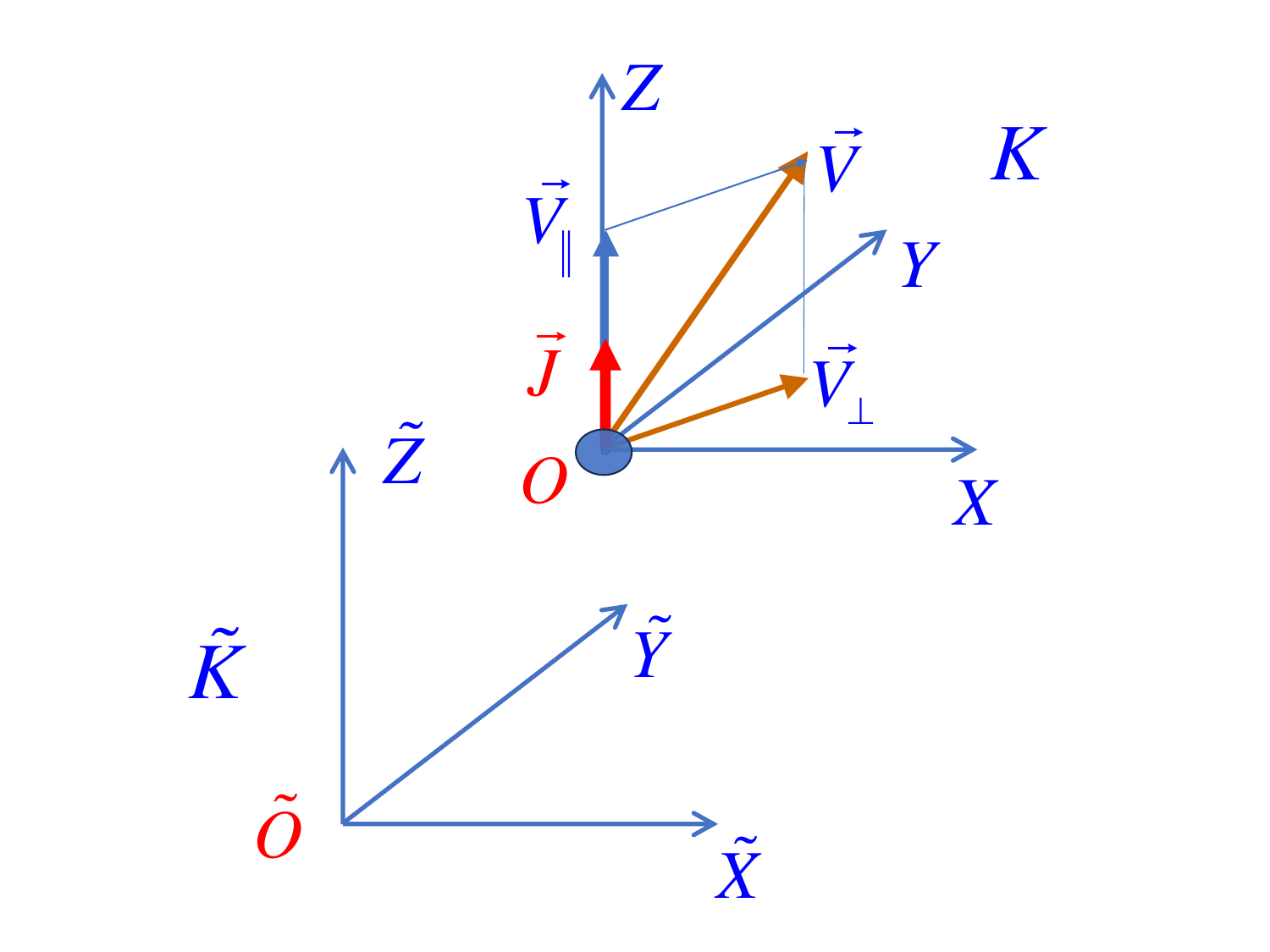}
    \caption{\n{F2}Two inertial frames, $K$ and $\tilde{K}$, are schematically shown at this figure. The Cartesian coordinates in these frames are $(T,X,Y,Z)$ and $(\tilde{T},\tilde{X},\tilde{Y},\tilde{Z})$, respectively.
    The frame $\tilde{K}$ is the "rest frame" of the homogeneous scalar field, in which it has the form $\Psi=\Psi_0 \tilde{T}$. The frame $K$ is the rest frame of the rotating black hole. The origin of this frame $O$ is at the position of the black hole. The frame $K$ moves  with respect to $\tilde{K}$ with the velocity $\vec{V}$. The coordinate axes in both frames are parallel. $Z$ axis in $K$ frame coincides with the direction of the spin $\vec{J}$ of the black hole. The velocity $\vec{V}$ in the frame $K$ can be decomposed as follows $\vec{V}=\vec{V}_{\parallel}+\vec{V}_{\perp}$, where $\vec{V}_{\parallel}$ is parallel to the spin of the black hole, and $\vec{V}_{\perp}$ is orthogonal to it.
    }
\end{figure}

Consider a rotating black hole moving with respect to $\tilde{K}$ with a constant velocity $\vec{V}$ (see figure~\ref{F2}).
We denote by $K$ a reference frame associated with the black hole and
denote by $X^{\mu}=(T,X,Y,Z)$  the Cartesian coordinates associated with this frame.
We choose the coordinate axes in the both frames to be in the same directions.
We also choose $Z$ axis to be parallel to the spin of the black hole.The vector $\vec{V}$ has the following components
\be
\begin{split}
&\vec{V}=(V_X,V_Y,V_Z)\, ,\\
& V^2=(\vec{V})^2=V_X^2+V_Y^2+V_Z^2\, .
\end{split}
\ee
Denote by $\vec{R}=(X,Y,Z)$ a 3D vector connecting the origin $\tilde{O}$ of the frame $\tilde{K}$ with the origin $O$ of the moving frame $K$. Then the Lorentz transformation implies
\begin{equation}
\begin{split}\n{LOR}
\tilde{T}&=\gamma\left( T+(\vec{V},\vec{R})\right) \, ,\\
\gamma&=1/\sqrt{1-V^2}\, .
\end{split}
\ee
The solution \eqref{PKP} written in the frame $K$ comoving with the black hole takes the form
\be \n{PSIV}
\begin{split}
\Psi=&\bar{\Psi}_0 (T+V_X X+V_Y  Y +V_Z Z)\, ,\\
 \bar{\Psi}_0 =&\dfrac{\Psi_0}{\sqrt{1-V^2}}\, .
\end{split}
\ee

\subsection{Scalar field solution in the presence of a moving rotating black hole}

In the previous subsection we ignore the gravitational field of the black hole. To obtain a solution for the scalar field in the presence of a moving rotating black hole we proceed as follows. We use the Kerr-Schild form of the metric \eqref{KSP} associated with $K$ frame, in which the black hole is at rest. In this form the Einstein equations are linearized. One can identify the metric $dS^2$ with the flat background geometry in $K$ frame, while the $\Phi l_{\mu}l_{\nu}$ describes its "perturbation" due to the presence of the black hole. We are looking for a solution of the scalar field equation \eqref{BOX}
which is regular at the horizon of the black hole and at far distance has the asymptotic form \eqref{PSIV}.

The required solution satisfying the imposed boundary conditions
in the  coordinates $(T,r,y,\phi)$ is
\be \n{SOL}
\begin{split}
&\Psi=\bar{\Psi}_0\left( T+T_0(r)+V_X \CAL{V}_X+V_Y \CAL{V}_Y+V_Z\CAL{V}_Z\right)\, ,\\
&T_0(r)=-2M\ln(r-r_-)\, ,\\
&\CAL{V}_X=\dfrac{1}{a}\sqrt{\dfrac{\Dy}{\Dr}}\left( (\Dr-Mr) \cos\phi -Ma\sin\phi\right)\, ,\\
&\CAL{V}_Y=\dfrac{1}{a}\sqrt{\dfrac{\Dy}{\Dr}}\left( (\Dr-Mr) \sin\phi +Ma\cos\phi\right)\, ,\\
&\CAL{V}_Z=\dfrac{y}{a}(r-M)\,  .
\end{split}
\ee
To check the validity of the boundary conditions at the infinity it is sufficient to use following asymptotics of the functions $\CAL{V}$
\be
\begin{split}
&\CAL{V}_X\approx r\sin\theta\cos\phi=X\, ,\\
&\CAL{V}_Y\approx r\sin\theta\sin\phi=Y\, ,\\
&\CAL{V}_Z\approx r\cos\theta=Z\, .
\end{split}
\ee
Since the coordinates $(T,r,y,\phi)$ are regular on the future horizon, and the components of its gradient $\Phi_{;\mu}$  are regular functions of these coordinates, one can conclude that the presented solution \eqref{SOL} does satisfy the required condition of the regularity at the horizon.

The same solution written in the Boyer-Lindquist coordinates $(t,r,\theta,\varphi)$   is
\be
\begin{split}
&\Psi=\bar{\Psi}_0\big( t+\tilde{t}_0+V_X \CAL{U}_X+V_Y \CAL{U}_Y+V_Z \CAL{U}_Z\big)\, ,\\
&\CAL{U}_X= \dfrac{\sin\theta}{\sqrt{\Dr}}\left[ (\Dr-Mr)\cos\psi-Ma\sin\psi\right]\, ,\\
&\CAL{U}_Y= \dfrac{\sin\theta}{\sqrt{\Dr}}\left[ (\Dr-Mr)\sin\psi+Ma\cos\psi\right]\, ,\\
&\CAL{U}_Z=V(r-M) \cos\theta\, ,\\
&\tilde{t}_0=t_0(r)+T_0(r)=\dfrac{M}{\sqrt{M^2-a^2}}\big[ r_+\ln(r-r_+)\\
&\quad -r_-\ln(r-r_-)\big]  -2M\ln(r-r_-)\, ,\\
&\varphi_0(r)=\dfrac{a}{2\sqrt{M^2-a^2}}\ln\big( \dfrac{r-r_+}{r-r_-}\big)-\arctan(r/a)
\, ,\\
&\psi=\varphi+\varphi_0(r)\, .
\end{split}
\ee

\section{Fluxes}

\subsection{Energy and angular momentum fluxes through  $r=$const surface}

Let us calculate the fluxes of the energy, angular momentum, and the momentum  through a 2D surface of constant radius $r=r_0$ surrounding the black hole. For these calculations it is convenient to use the Boyer-Lindquist coordinates $(t,r,\theta,\varphi)$. Denote by $\Sigma_0$ a 3D timelike surface describing the "evolution" of $S$ for the time interval $(t_-,t_+)$.

Denote by $\ts{q}$ a 3D metric on $\Sigma$ induced by its embedding in the 4D space. Then
\be
q\equiv \sqrt{-\mbox{det}(\ts{q})}=\sqrt{\Delta\Sigma}\sin\theta\, .
\ee
A unit vector $\ts{n}$ orthogonal to $\Sigma_0$ and inward directed is
\be
n^{\mu}=-\sqrt{\Delta/\Sigma}\delta^{\mu}_r\, .
\ee
The vector of the volume element of the surface $\Sigma$ is
\be
d\sigma^{\mu}=n^{\mu} dt\, d\theta\, d\varphi=-\delta_r^{\mu}
\Delta \sin\theta\, dt\, d\theta \, d\varphi\, .
\ee

We denote the fluxes of the energy and angular momentum through 3D surface $\Sigma_0$ into its interior  per a unit time $t$ by $\EE$ and $\JJ$, respectively. Then one has
\be
\begin{split} \n{E_FLUX}
\EE=&-\dfrac{1}{t_+-t_-}\int_{\Sigma} \xi^{\mu}T_{\mu\nu} d\sigma^{\mu}=\Delta \int_S T_{tr} d\omega\, ,\\
\JJ=&\dfrac{1}{t_+-t_-}\int_{\Sigma} \zeta^{\mu}T_{\mu\nu} d\sigma^{\mu}=-\Delta \int_S T_{r\varphi} d\omega\, ,\\
&d\omega=\sin\theta d\theta\, d\varphi \, ,\\
\end{split}
\ee
 The signs in these expressions are chosen so that these quantities describe the flux into the surface $S$ from its exterior.

Since $g_{tr}$ and $g_{r\varphi}$ components of the Kerr metric in the Boyer-Lindquist coordinates vanish one has
\be
T_{tr}=\Psi_t \Psi_r\hh T_{t\varphi}=\Psi_r \Psi_{\varphi}\, .
\ee
Calculating these expressions and taking integrals in \eqref{E_FLUX} one obtains
\be
\begin{split} \n{FLUXES}
&\EE=8\pi \bar{\Psi}_0^2 M r_+\, ,\\
&\JJ=- \dfrac{4}{3}\pi \bar{\Psi}_0^2 a M^2 (V_X^2+V_Y^2)\, .
\end{split}
\ee
Let us emphasize that  these quantities do not depent on the radius $r=r_0$ of the surface $S$ which was used for their calculation. (For explanation of this property and more details, see appendix~B.)

\subsection{Momentum fluxes}

We use the vectors $e_{i}^{\mu}$, defined in \eqref{eee} to define the following objects\footnote{
Let us note that the norm of the vectors $e_{i}^{\mu}$, calculated in the metric $g_{\mu\nu}$, at far distance slightly differs from one. However, it is possible to check that using the normalized versions of these vectors in the expressions given below does not change the results when the limit $r=r_0\to\infty$ is taken.
}
\be
\begin{split} \n{XYZ_FLUX}
\PP_X=&\dfrac{1}{t_+-t_-}\int_{\Sigma} T_{\mu\nu}e_{(X)}^{\mu} d\sigma^{\nu}=-\Delta \int_S T_{X r}d\omega \, ,\\
\PP_Y=&\dfrac{1}{t_+-t_-}\int_{\Sigma} T_{\mu\nu}e_{(Y)}^{\mu} d\sigma^{\nu}=-\Delta \int_S T_{Y r}d\omega \, ,\\
\PP_Z=&\dfrac{1}{t_+-t_-}\int_{\Sigma}  T_{\mu\nu}e_{(Z)}^{\mu} d\sigma^{\nu}=-\Delta \int_S T_{Z r}d\omega \, ,\\
T_{X r}=& T_{\mu r}e_{(X)}^{\mu}\hhh  T_{Y r}= T_{\mu r}e_{(Y)}^{\mu}\hhh  T_{Z r}= T_{\mu r}e_{(Z)}^{\mu}\, .
\end{split}
\ee
Calculating these integrals and taking the limit $r_0\to\infty$ one obtains the following results
\be
\begin{split} \n{PXYZ}
&\PP_X=-8\pi\bar{\Psi}_0^2 \left[  M r_+ V_X +\dfrac{2}{3} a M V_Y\right] \, ,\\
&\PP_Y=-8\pi\bar{\Psi}_0^2 \left[M r_+ V_Y-\dfrac{2}{3} a M V_X \right]\, ,\\
&\PP_Z=-8\pi \bar{\Psi}_0^2 M r_+ V_Z\, .
\end{split}
\ee

The obtained results can be presented in the following 3D vector  form. Let us denote
\be
\begin{split}
&\beta=8\pi {\Psi}_0^2\hh \vec{V}=(V_X,V_Y,V_Z)\, ,\\
& \vec{\PP}=(\PP_X,\PP_Y,\PP_Z)\hhh  \vec{J}=M\vec{a}\hhh \vec{\JJ}=\dfrac{\vec{a}}{a}\JJ
\, .
\end{split}
\ee
Here $\vec{a}$ is a 3D vector with the norm $a$ and directed along $Z-$axis. Consider an observer which is located at very far  distance from the black and assume that the black hole is initially at rest in his/her frame. As a result of the interaction with "moving" scalar field, the black hole absorbs energy and momentum. To describe its further evolution as a result of this effect, a far-distant observer can neglect the black hole size and approximate it by the massive point with spin which has energy and momentum. In this interpretation the calculated quantities $\EE$ and $\vec{\PP}$ are nothing but the components of the 4D force $\ts{F}$ acting on such a massive point.

In the asymptotic flat coordinates $(T,X,Y,Z)$ the components of this 4D force are
\be\n{FK}
 \begin{split}
&F^{\mu}=(\EE,\PP_X,\PP_Y,\PP_Z)\, ,\\
&F^T=\dfrac{\beta M r_+}{1-V^2}\, ,\\
&\vec{F}=-\dfrac{\beta}{1-V^2}\left(M r_+ \vec{V}+\dfrac{2}{3} \vec{J}\times \vec{V}\right) \, .
\end{split}
\ee
In addition to these expressions for the 4D force acting on the black hole, there exists one more equation which demonstrates that the spin of the black changes when the black hole has a non-vanishing transverse component of the velocity $V_{\perp}$
\be\n{JK}
 \begin{split}
& \vec{\JJ} =-\dfrac{1}{6}\beta M V_{\perp}^2 \vec{J}\, ,\\
& V_{\perp}^2=\vec{V}^2 -\dfrac{1}{J^2} (\vec{J}\cdot \vec{V})^2\, .
\end{split}
\ee

\section{Motion of a rotating black hole through the scalar field}

\subsection{Friction force in $\widetilde{K}$ frame}

The 4D force $F^{\mu}$ is calculated in the frame where the black hole is initially at rest. Under the action of this force the black hole has a non-vanishing acceleration. Since the velocity of the black hole interacting with the scalar fields changes in time, the  frame in which  the black hole is at rest is not inertial. For this reason it is more convenient to write the equations of motion in the frame $\tilde{K}$ associated with the scalar field. Let us denote by $f^{\mu}\equiv \tilde{F}^{\mu}$ the components of the 4D force in $\tilde{K}$ frame. The corresponding Lorentz transformation implies
\be
 \begin{split}
&f^0=\gamma (F^T+\vec{V}\cdot\vec{F})\, ,\\
&\vec{f}=\vec{F}+\gamma F^T \vec{V}+(\gamma-1)\dfrac{\vec{V}\cdot\vec{F}}{V^2}\vec{V}\, .
\end{split}
\ee
Here  $\gamma=1/\sqrt{1-V^2}$.
Using \eqref{FK} one obtains
\be
 \begin{split}\n{fff}
&f^0=\dfrac{\beta M r_+}{\sqrt{1-V^2}}\, ,\\
&\vec{f}=-\dfrac{2}{3} \dfrac{\beta}{\sqrt{1-V^2}}\ [ \vec{J}\times\vec{V}]\, .
\end{split}
\ee
Let us notice that the following two useful relations are valid
\be
\vec{V}\cdot \vec{f}=0\hh \vec{J}\cdot \vec{f}=0\, .
\ee

\subsection{Equations of motion}

We denote  by $\ts{P}$ the 4D momentum of the black hole in $\tilde{K}$ frame. In Cartesian coordinates it has the following form
\be
P^{\mu}=(E,\vec{P})\hh \vec{P}=(P_X,P_Y,P_Z)
\, .
\ee
The energy $E$ and momentum $\vec{P}$ of the black hole with mass $M$ and velocity $\vec{V}$ are
\be
E=\dfrac{M}{\sqrt{1-V^2}}\hh \vec{P}=\dfrac{M \vec{V}}{\sqrt{1-V^2}}\, .
\ee
Such a black hole is similar to a massive relativistic particle, with two important differences: (i) The mass of the black hole changes as a result of the absorption of the scalar field, and (ii) the black hole has spin $\vec{J}$, which also changes with time.

The equations of motion are
\be\n{ff}
\dfrac{dE}{d{\tau}}=f^0\hh \dfrac{d\vec{P}}{d{\tau}}=\vec{f}\, ,
\ee
where $f^0$ and $\vec{f}$ are given in \eqref{fff}, and $\tau$ is the proper time along the worldline of the black hole.
There exists an additional equation for the spin evolution
\be \n{JEE}
\dfrac{dJ}{d{\tau}}=-\JJ\, ,
\ee
where $\JJ$ is given by \eqref{FLUXES}.

Since vector $\vec{P}$ is collinear with $\vec{V}$, one has
\be
\dfrac{1}{2} \dfrac{d\vec{P}^2}{d\tau}=\vec{P}\cdot \dfrac{d\vec{P}}{d{\tau}}=\dfrac{M }{\sqrt{1-V^2}}\, \vec{V}\cdot \vec{f}=0\, .
\ee
Hence $\vec{P}^2 =P_0^2=$const.

Let us write the momentum vector in the form
\be
\vec{P}=\vec{P}_{\parallel}+\vec{P}_{\perp}\, ,
\ee
where $\vec{P}_{\parallel}$ is parallel to the spin vector $\vec{J}$ and $\vec{P}_{\perp}$ is perpendicular to it. Then one has
\be
\vec{P}_{\parallel}\cdot \dfrac{d\vec{P}_{\parallel}}{d{\tau}}=0\, .
\ee
This means that  the norms of the both vectors $P_{\parallel}=\sqrt{\vec{P}_{\parallel}^2}$ and $P_{\perp}=\sqrt{\vec{P}_{\perp}^2}$ are conserved quantities.

Using relations
\be
 \begin{split}
&E=\sqrt{M^2+P_0^2}\, ,\\
& \dfrac{1}{\sqrt{1-V^2}}=\dfrac{ 1}{M} \sqrt{M^2+P_0^2}\, ,
 \end{split}
 \ee
one can rewrite the first equation in \eqref{ff} as follows
\be \n{eqM}
 \begin{split}
&\dfrac{dM}{d\tau}=\beta \dfrac{r_+}{M} (M^2+P_0^2) \, ,\\
& r_+=M+\sqrt{M^2-J^2/M^2}
\, .
 \end{split}
 \ee

Using relation
\be
V_{\perp}^2=\dfrac{P_{\perp}^2}{M^2+P_0^2}
\ee
one can write \eqref{JEE} in the form
\be \n{eqJ}
\dfrac{dJ}{d\tau}=-\dfrac{1}{6} \beta \dfrac{M P_{\perp}^2}{M^2+P_0^2} J\, .
\ee

The second equation in \eqref{ff} implies
\be \n{perp}
\dfrac{d\vec{P}_{\perp}}{d\tau}=-\dfrac{2}{3} \beta \dfrac{1}{M} [\vec{J}\times \vec{P}_{\perp}]\, .
\ee
Let us denote
\be
P_X +i P_Y={P}_{\perp} \exp(i\alpha)\, .
\ee
Then the equation \eqref{perp} implies
\be \n{eqa}
\dfrac{d\alpha}{d\tau}=-\dfrac{2}{3}\dfrac{\beta J}{\sqrt{M^2+P_0^2}}\, .
\ee

The obtained equations \eqref{eqM} and \eqref{eqJ} allow one to find the time dependence of the black hole mass $M$ and spin $J$ for a given initial values $M_0$ and $J_0$ of these parameters.  The equation \eqref{eqa} determines the time evolution of the transverse momentum $\vec{P}_{\perp}$. In order to find a unique solution one needs also specify the conserved parameters $P_0$ and $P_{\perp}$.

Let us make a following remark. In the above consideration we assumed that a distant observer is located at far distance $L$ from the black hole, $L>>M$. This allows one to describe a black hole as a "point-like particle" which has mass $M$ and spin $J$. We also assume that these parameters change slowly in time, so that
\be \n{MMJJ}
\dot{M}/M\ll 1/M\hh \dot{J}/J\ll 1/M\, .
\ee
In this adiabatic approximation one can describe the black hole metric by the Kerr solution with a slowly changing parameters $M(t)$ and $J(t)$. Note that such an adiabatic approximation is broken in the vicinity of the moment of time, at which the mass of the black hole formally becomes infinite.

\subsection{General properties of solutions}

\subsubsection{Mass evolution}

Let us discuss  general properties of the solutions of the equations of motion for the rotating black hole moving in the homogeneous scalar field. First of all let us notice that equations \eqref{eqM} and \eqref{eqJ} show that the black hole mass $M$ is a monotonically increasing function of time $\tau$, while its spin $J$ monotonically decreases in time.

One can obtain a more detailed information about the time dependence of these parameters by using the following trick. Let us denote $r_+=b M$, where $1\le b\le 2$. The quantity $b$ takes the value 1 for the extremely rotating black hole when $J=M^2$, and $b=2$ for a nonrotating black hole.  Solving \eqref{eqM} for $b$=const one gets
\be
\arctan(M_b/P_0)-\arctan(M_0/P_0)=b\beta P_0\tau\, ,
\ee
where $M_0$ is the mass  at the initial moment of time $\tau=0$. If $M(\tau)$ is the exact solution of the equation \eqref{eqM} then for the same initial mass $M_0$ one has
\be
M_1(\tau)\le M(\tau)\le M_2(\tau)\, .
\ee
Denote
\be
\tau_0=\dfrac{1}{\beta P_0}\left(\dfrac{\pi}{2}--\arctan(M_0/P_0)\right)\, .
\ee
Then the black hole mass $M$ becomes infinite at the finite proper time $\tau=\tau_{fin}$
\be
\dfrac{1}{2}\tau_0\le \tau_{fin}\le \tau_0\, .
\ee

\subsubsection{Spin evolution}

Let us consider the spin of the black hole as a function of its mass, $J=J(M)$, Then using \eqref{eqM} and \eqref{eqJ} one obtains
\be \n{JMexact}
\dfrac{1}{J}\dfrac{dJ}{dM}=-\dfrac{1}{6}\dfrac{P_{\perp}^2 M^2}{r_+ (M^2+P_0^2)^2}\, .
\ee
Substituting $r_+=bM$ and solving the obtained equation with fixed value of $b$ one finds
\be
J_b(M)=J_0 \exp\big[-\dfrac{1}{6b}\dfrac{P_{\perp}^2(M^2-M_0^2)}{(M^2+P_0^2)(M_0^2+P_0^2)}\big]\, .
\ee
The  exact solution $J(M)$ of \eqref{JMexact} satisfies the inequalities
\be
J_1(M)\le J(M)\le J_2(M)\, .
\ee
For $M\to \infty$
\be
J_{b,fin}=J_0 \exp\big[-\dfrac{P_{\perp}^2}{6b(M_0^2+P_0^2)}\big] \, .
\ee
Hence at the moment when the mass of the black hole becomes infinitely large, its spin remains finite $J=J_{fin}$
\be
J_{1,fin}\le J_{fin}\le J_{2,fin}\, .
\ee

\section{Special cases}

\subsection{Black hole at rest in $\widetilde{K}$ frame}

If the black hole is at rest then $P_0=0$ and the spin of the black hole remains constant.  The equation for the black hole mass takes the form
\be \n{MP0}
\dfrac{dM}{d\tau}=\beta M r_+=\beta \big( M^2+\sqrt{M^4-J^2}\big) \, .
\ee
If the spin $J$ does not vanish one can denote
\be
\mu=M/\sqrt{J}\, ,
\ee
and write a solution of the equation \eqref{MP0} in the form
\be
\sqrt{J}\beta\tau=\hat{\tau}\hh
\hat{\tau}=\int \dfrac{d\mu}{\mu^2+\sqrt{\mu^4-1}}\, .
\ee
Taking the integral one obtains
\be
\hat{\tau}=\dfrac{1}{3}\big[ \mu^3-1-\mu\sqrt{\mu^4-1}+2(F(i\mu,i)-F(i,i))\big]\, .
\ee
Here $F(a,b)$ is the  incomplete elliptic integral of the first kind, and
the integration constant in this expression is chosen so that $\hat{\tau}=0$ for $\mu=1$.
For $\mu\to\infty$ the parameter $\hat{\tau}$ has a limit $0.54068$.
The plot of $\hat{\tau}$ as a function of $\mu$ is shown at figure~\ref{F3}.
\begin{figure}[!hbt]
    \centering
\includegraphics[width=0.35\textwidth]{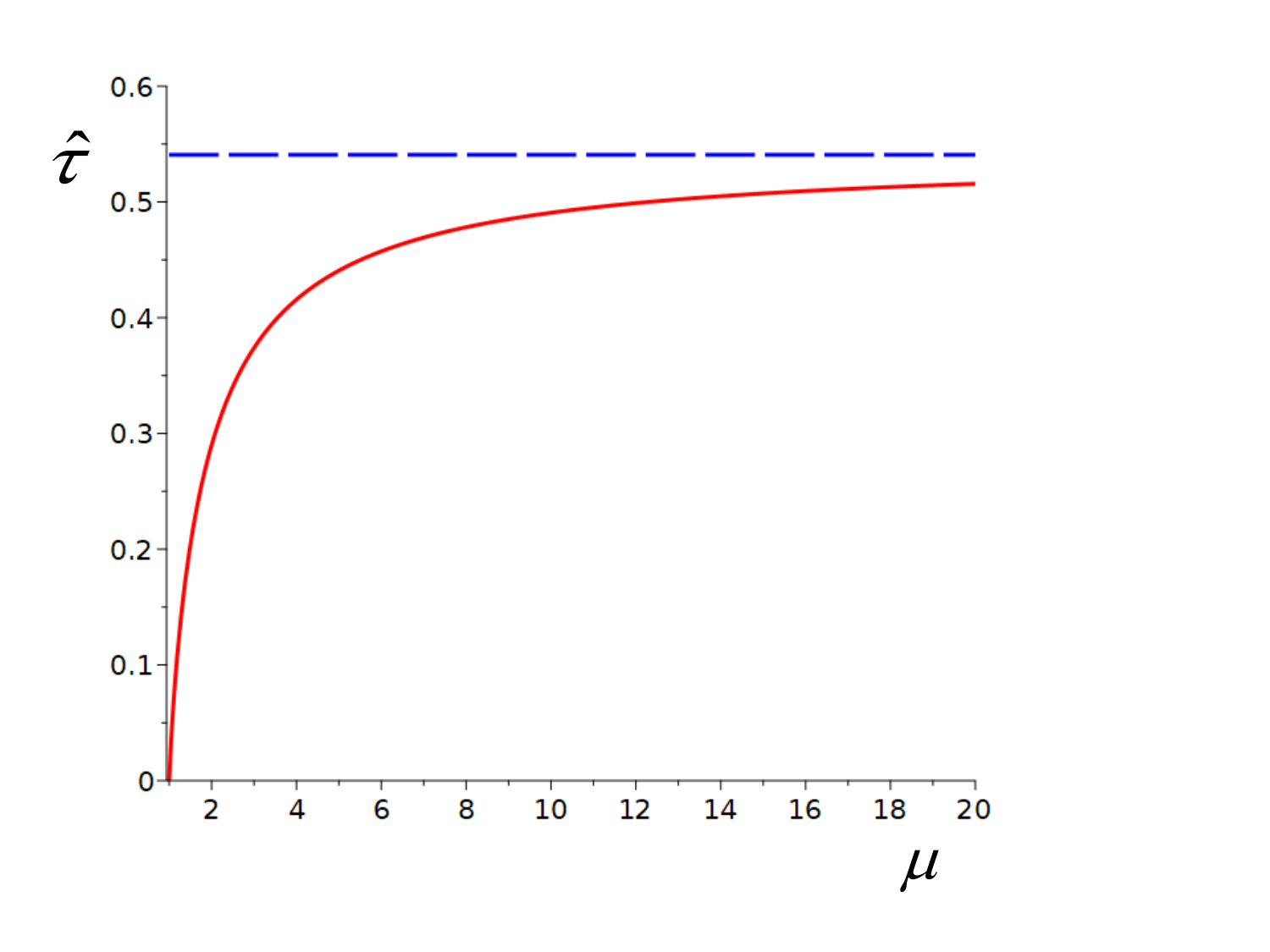}
    \caption{\n{F3} $\hat{\tau}$ as a function of $\mu$. Dashed line shows the limiting value of $\hat{\tau}$ for $\mu\to\infty$, which is equal to 0.54068.  }
\end{figure}

If $J=0$ one obtains the following solution of the equation \eqref{MP0} for the mass
\be
M=\dfrac{M_0}{1-2\beta M_0\tau}\, .
\ee

\subsection{Motion of a nonrotating black hole}

When the spin of the black hole vanishes, one can always chose the orientation of the axes so that $P_{\perp}=0$ and $P_{\parallel}=P_0$. One also has $r_+=2M$.  The equation \eqref{eqM} simplifies and takes the form
\be
\dfrac{dM}{d\tau}=2\beta (M^2+P_0^2)\, .
\ee
It can be easily integrated with the following result
\be
2\beta P_0\tau=\arctan(M/P_0)\, .
\ee
This expression correctly reproduces the result obtained in \cite{FFM}.

\subsection{Motion in the spin direction}

For the motion in the spin direction one has $P_{\perp}=0$ and, as a result,  the spin of the black hole remains constant. If its initial value does not vanish, we denote
\be
\mu=M/\sqrt{J}\hh p=P_0/\sqrt{J}\, .
\ee
Then one has
\be
\beta \sqrt{J}\tau=\int \dfrac{\mu^2 d\mu}{(\mu^2+p^2) (\mu^2+\sqrt{\mu^4-1})}\, .
\ee
The integral in the right-hand side of this relation can be expressed in terms of the incomplete elliptic integrals.

\subsection{Transverse to the spin motion}

For this case $P_{\parallel}=0$ and $P_0=P_{\perp}$. We denote
\be
M=P_{\perp} m\hh J=P_{\perp} j\hh \hat{\tau}=\beta P_{\perp}\tau\, .
\ee
Then the equations for the mass and spin evolution take the form
\be \n{eqm}
 \begin{split}
&\dfrac{dm}{d\hat{\tau}}= \dfrac{\hat{r}_+(m^2+1)}{m}  \, ,\\
& \dfrac{dj}{d\hat{\tau}}= -\dfrac{1}{6}\dfrac{m j}{m^2+1}\, ,\\
&\hat{r}_+=m+\sqrt{m^2-j^2/m^2}
\, .
 \end{split}
 \ee

\section{Discussion}

In this paper we discussed a motion of a rotating black hole in the homogeneous  massless scalar field.
For this purpose we used the Kerr-Schild form of the Kerr metric. We introduced  two frames. One which we denoted by $\tilde{K}$ is the frame in which the asymptotic scalar field does not depend on the spatial coordinates. The other frame, which we denoted by $K$, is a frame moving with respect to $\tilde{K}$ with a constant velocity $V$ and in which the black hole is at rest. We first solved the scalar field equation in $K$ frame and found a solutions
satisfying the condition of the regularity at the black hole horizon and proper behavior at the infinity. After this we calculated the fluxes of the  energy, momentum and angular  momentum through  a surface surrounding the black hole. This allowed us to find the force acting on the black hole and to obtain the equation of its motion in $\tilde{K}$ frame.

The main results of the analysis of the solutions of these equations are the following:
\begin{itemize}
\item For a general type of motion the components of the momentum of the black hole both in the direction of its spin $P_{\parallel}$ and in the transverse plane  $P_{\perp}$ are conserved.
\item The black hole mass $M$ monotonically grows and formally becomes infinitely large in a finite interval of time $\tau_{fin}$, which depends both on the strength of the scalar field and initial value of the mass, spin and velocity of the black hole.
\item The spin $J$ of the black hole monotonically decrease, but does not vanish at $\tau=\tau_{fin}$. However, since the mass of the black hole becomes infinitely large, the rotation parameter $s=J/M$ vanishes in this limit, and the Kerr metric reduces to its Schwarzschild limit.
\end{itemize}
Let us note that these results are obtained in the adiabatic approximation in which the mass $M$ and the spin $J$ of the black hole change slowly in time. At time close to $\tau_{fin}$, where $\dot{M}$ is not small, this approximation is broken.

In this paper we considered a simple model and  assumed that the scalar field  obeys a linear equation $\Box{\Phi}=0$. This approximation can be violated in application to some interesting cases, e.g. for the scalar field of the ghost condensate, where the effects of the non-linearity might become important. In application to the cosmology, the scalar field may depend on the general expansion of the universe, which would affect the imposed boundary conditions at the infinity. For the black hole in the expanding universe the adopted form \eqref{KS} of the metric should also be modified. It is  interesting to study how these modification of the model would affect the interaction of the black hole with a scalar field and its motion.

\appendix

\section{Complex null tetrads}

In this appendix we collect useful formulas and expressions for the complex null tetrads in the Kerr-Schild geometry.
Let us denote
\begin{equation}
\begin{split}
m^{\mu}&=\sqrt{\dfrac{\Dy}{2\Sigma}}\left( 1,0,i,\dfrac{a}{\Dy}\right)\, ,\\
\bar{m}^{\mu}&=\sqrt{\dfrac{\Dy}{2\Sigma}}\left( 1,0,-i,\dfrac{a}{\Dy}\right)\, .
\end{split}
\end{equation}
These vectors satisfy the following relations
\begin{equation}
\begin{split}
&g_{\mu\nu}m^{\mu}m^{\nu}=\eta_{\mu\nu}m^{\mu}m^{\nu}=0\, ,\\
&g_{\mu\nu}\bar{m}^{\mu}\bar{m}^{\nu}=\eta_{\mu\nu}\bar{m}^{\mu}\bar{m}^{\nu}=0\, ,\\
&g_{\mu\nu}m^{\mu}\bar{m}^{\nu}=\eta_{\mu\nu}m^{\mu}\bar{m}^{\nu}=1\, .
\end{split}
\end{equation}
The complex null vectors $\ts{m}$ and $\bar{\ts{m}}$ are orthogonal to $\ts{l}$ both in $\ts{g}$ and $\ts{\eta}$ metrics.

The forth vector of the null tetrad has a slightly different form for  $\ts{g}$ and $\ts{\eta}$ metrics.
We denote
\be
K^{\mu}=\dfrac{\Dr}{2\Sigma}\left(1,1,0,\dfrac{a}{r^2+a^2}\right)\, .
\ee
This vector is null in the metric $\ts{\eta}$ and normalized so that
\be
\eta_{\mu\nu}l^{\mu}K^{\nu}=-1\, .
\ee
A similar vector $\ts{k}$ for $\ts{g}$ metric is
\be
k^{\mu}=K^{\mu}+\dfrac{1}{2}\Phi l^{\mu}\, .
\ee
It satisfies the relations
\be
\begin{split}
&g_{\mu\nu}k^{\mu}k^{\nu}=0\hh g_{\mu\nu}l^{\mu}k^{\nu}=-1\, ,\\
&g_{\mu\nu}m^{\mu}k^{\nu}= g_{\mu\nu}\bar{m}^{\mu}k^{\nu}=0\, .
\end{split}
\ee
The complex null tetrad in the metric $\ts{g}$ regular at the future horizon of the Kerr black hole is
\be\n{TETRAD}
\ts{z}^{a}=(\ts{l},\ts{k},\ts{m},\bar{\ts{m}})\, .
\ee
The index $a$ enumerating the basis vectors takes the values $0,1,2,3$.
By the construction the vectors of the basis $\ts{z}^{a}$ are regular the future event horizon.

\section{Energy and angular momentum fluxes through the horizon}

\n{EJF}

Let $\xi^{\mu}$ be a Killing vector and $T_{\mu\nu}$  be a conserved stress-energy tensor, $T^{\mu\nu}_{\ \ ;\nu}=0$. Then the following vector
\be
\CAL{P}^{\mu}=T^{\mu\nu} \xi_{\nu}\, .
\ee
is conserved
\be
\CAL{P}^{\mu}_{\ ;\mu}=0\, .
\ee

\begin{figure}[!hbt]
    \centering
\includegraphics[width=0.5\textwidth]{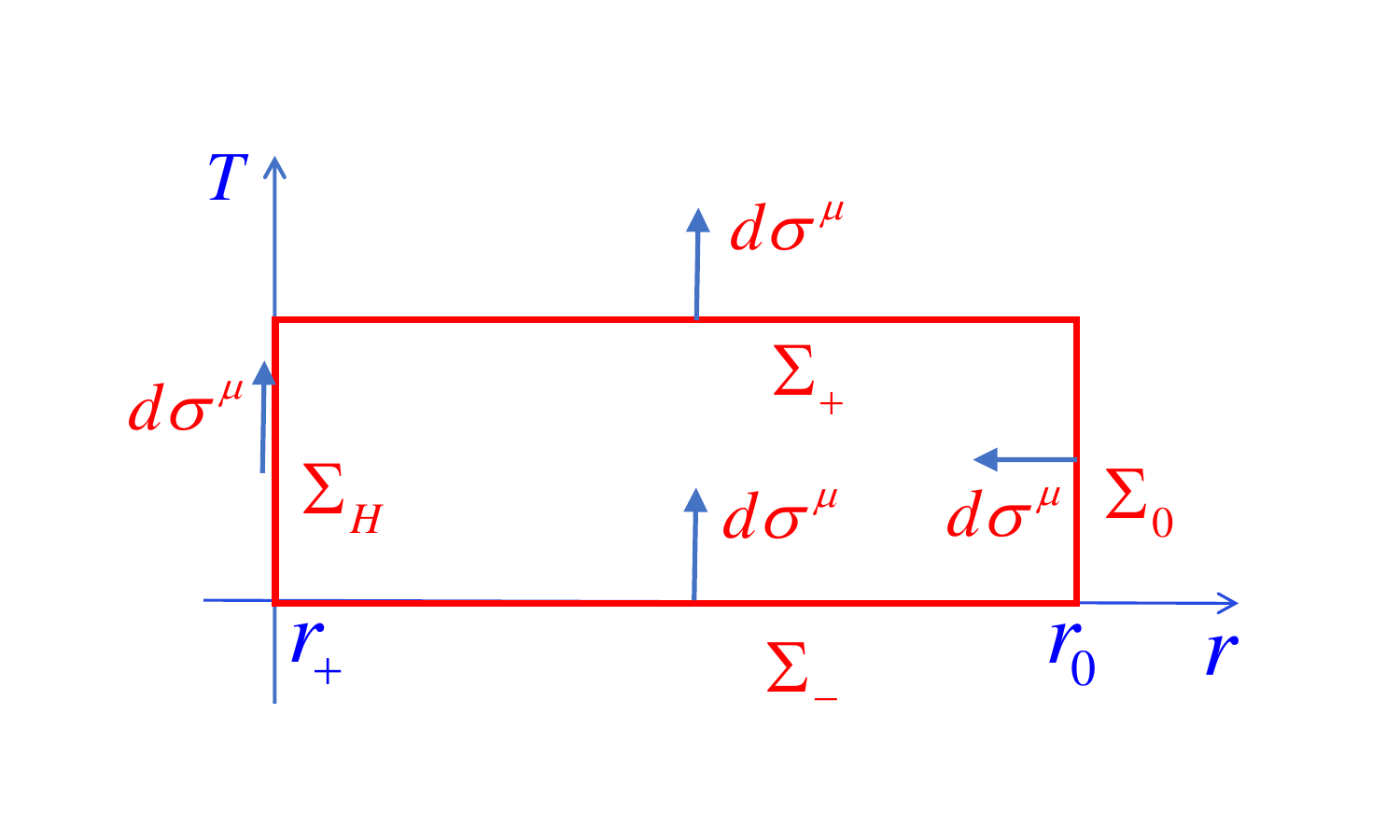}
    \caption{\n{F4} Illustration to the Stockes' theorem.}
\end{figure}

Let us consider a rotating black hole and use the coordinates $(T,r,y,\phi)$. We denote by $\Sigma_{\pm}$ two 3D slices determined by the equations $T=T_{\pm}$ and restricted from one side by the horizon $r=r_+$ and from the other side by a surface $r=r_0>r_+$. We denote by $\Sigma_H$ a part of the  horizon surface between $T=T_-$ and $T_+$. Similarly, we denote be $\Sigma_0$ a 3D surface $r=r_0$ between $T=T_-$ and $T_+$ (see figure~\ref{F4}).

Let $\CAL{V}$ be a four volume restricted by $\Sigma_{\pm}$, $\Sigma_H$ and $\Sigma_0$. Using the Stockes' theorem one can write
\be
\begin{split} \n{SIG}
0=&  \int_{\CAL{V}} \sqrt{-g}d^4 x \CAL{P}^{\mu}_{\ ;\mu}=\\
=& \big[ \int_{\Sigma_+}-\int_{\Sigma_-} \big] \CAL{P}^{\mu} d\sigma_{\mu}+
 \big[ \int_{\Sigma_{0}}-\int_{\Sigma_H}  \big] \CAL{P}^{\mu} d\sigma_{\mu}
\, .
\end{split}
\ee
The surface elements $d\sigma^{\mu}$ are chosen so that for $\Sigma_{\pm}$ and at the horizon $\Sigma_H$ they are both future directed, while at $\Sigma_0$ it is directed into this surface's interior.

For the problem under the consideration the gradient of the scalar field $\Psi$ does not depend on time, and, since the metric is also time independent, the stress-energy tensor has the same property. As a result
the expression in the first square bracket in \eqref{SIG} vanishes and one has
 \be \n{CONS}
 \int_{\Sigma_{0}}  \CAL{P}^{\mu} d\sigma_{\mu}=   \int_{\Sigma_H}   \CAL{P}^{\mu} d\sigma_{\mu}\, .
 \ee
This relation shows that
\begin{itemize}
\item The flux of $\ts{\CAL{P}}$ inside $\Sigma_0$ during  the time interval $T_+-T_-$ is equal to the flux through the horizon for the same interval of time $T$.
\item The flux of $\ts{\CAL{P}}$ inside $\Sigma_0$ in fact does not depend in the choice of the radius $r_0$.
\end{itemize}
Let us emphasize that at the surface of constant radius $r=r_0$ the coordinates $T$, $v$, and $t$ differs only by constant values. For this reason, one has
\be
T_+-T_-=v_+-v_-=t_+-t_-\, .
\ee
These remarks can be used to confirm the results \eqref{EEE}  for the energy and angular momentum fluxes through $\Sigma_0$.

Let us calculate the energy and angular momentum fluxes through the horizon of the black hole. Denote by
\be
\eta^{\mu}=\xi^{\mu}+\Omega \zeta^{\mu}\hh \Omega=\dfrac{a}{2Mr_+}\, .
\ee
Then for the horizon surface $\Sigma_H$ one has
\be
d\sigma^{\mu}=-\eta^{\mu} \dfrac{2Mr_+}{a} dy\, d\phi\, dT\, .
\ee

The energy and angular momentum fluxes through the horizon are
\be
\begin{split} \n{AEEE}
\EE=&-\dfrac{1}{T_+-T_-}\dfrac{2Mr_+}{a}\int_{\Sigma_H} \xi^{\mu}\eta^{\nu}T_{\mu\nu} dy\, d\phi\, dT\\
=& \dfrac{2 M r_+}{a} \int_{-a}^{a} (T_{TT}+\Omega T_{T\phi}) dy\, d\phi\, ,\\
\JJ=&\dfrac{1}{T_+-T_-}\dfrac{2Mr_+}{a}\int_{\Sigma_H} \zeta^{\mu}\eta^{\nu}T_{\mu\nu}dy\, d\phi\, dT\\
=& -\dfrac{2 M r_+}{a} \int_{-a}^{a} (T_{T\phi}+\Omega T_{\phi\phi}) dy\, d\phi\, .
\end{split}
\ee
Calculations give
\be
\begin{split}
&T_{TT}+\Omega T_{T\phi} \\
&\eqH \bar{\Psi}^2_0\big[
1-\dfrac{(r_-\cos\phi+a\sin\phi)\sqrt{a^2-y^2}}{2\sqrt{2Mr_+} a}V_X\\
&-\dfrac{(r_-\sin\phi-a\cos\phi)\sqrt{a^2-y^2}}{2\sqrt{2Mr_+} a}V_Y
\big]
\, .
\end{split}
\ee
After integration of this expression over the angle $\phi$ the terms which depend on the velocity components $V_X$ and $V_Y$ vanish. The further integration over $y$ gives
\be
\EE\eqH 8\pi  \bar{\Psi}^2_0 M r_+ \, .
\ee
This expression correctly reproduces the result \eqref{FLUXES} as it should be.

To calculate the angular momentum flux through the horizon one need first to find
the value of $T_{T\phi}+\Omega T_{\phi\phi}$  on the horizon. The corresponding expression is rather long and we do not reproduce it here. Instead of this, we give the expression which is obtained after integration of this object over the angle $\phi$
\be
\int_0^{2\pi} (T_{T\phi}+\Omega T_{\phi\phi}) d\phi=- \bar{\Psi}^2\dfrac{\pi M}{2ar_+} (V_X^2+V_Y^2) (a^2-y^2)\, .
  \ee
Using \eqref{AEEE} and performing the integration  over $y$  one obtains
\be
\JJ\eqH - \dfrac{4}{3}\pi \bar{\Psi}_0^2 a M^2 (V_X^2+V_Y^2)\, .
\ee
This result correctly reproduces the expression obtained earlier in \eqref{FLUXES}.
The calculations presented in this appendix provide an additional check of the results presented in section~IV.

\acknowledgments

The author would like to thank Shinji Mukohyama and Andrei Frolov for stimulating discussions.
This work was supported  by  the Natural Sciences and Engineering
Research Council of Canada. The  author is also grateful to the
Killam Trust for its financial support.


%

\end{document}